\documentclass[11pt]{article}
\usepackage[margin=1in]{geometry}
\usepackage{amsmath,amsfonts,amsthm,mathrsfs,mathpazo,xspace,hyperref}

\newtheorem{prop}{Proposition}[section]

\newtheorem{theorem}[prop]{Theorem}
\newtheorem{fact}[prop]{Fact}

\newtheorem{lemma}[prop]{Lemma}

\newtheorem{corollary}[prop]{Corollary}
\newtheorem{claim}[prop]{Claim}
\theoremstyle{definition}
\newtheorem{definition}[prop]{Definition}
\newtheorem*{remark}{Remark}

\usepackage{color}
\definecolor{mygrey}{gray}{0.50}

\newcommand{\mypar}[1]{\paragraph{#1.}}

\newcommand{\RR}{\mathbb{R}}
\renewcommand{\SS}{\mathbb{S}}

\newcommand{\cX}{\mathcal{X}}
\newcommand{\cY}{\mathcal{Y}}

\newcommand\eps\varepsilon
\renewcommand\b{\{0,1\}}
\newcommand\bstar{\{0,1,\star\}}
\newcommand{\rrho}{p}

\newcommand{\ceq}{\subseteq}

\DeclareMathOperator{\cost}{cost}
\DeclareMathOperator{\err}{err}
\DeclareMathOperator{\D}{D}
\DeclareMathOperator{\Exp}{\mathbb{E}}
\DeclareMathOperator*{\spn}{span}
\DeclareMathOperator{\R}{R}

\newcommand{\dist}{\Delta}

\newcommand{\disj}{\textsc{disj}}
\newcommand{\ghd}{\textsc{ghd}}

\title{\bf
An Optimal Lower Bound on the Communication Complexity of
\textsc{Gap-Hamming-Distance}}
\author{
  Amit Chakrabarti\footnote{Department of Computer Science, Dartmouth College, Hanover, NH 03755, USA.
  Supported by NSF Grant IIS-0916565 and a McLane Family Fellowship.}
\and
  Oded Regev\footnote{CNRS, D{\'e}partement d'Informatique, {\'E}cole normale sup{\'e}rieure, Paris and Blavatnik School of Computer Science, Tel Aviv University. Supported by the Israel Science Foundation, by the Wolfson Family
  Charitable Trust, and by a European Research Council (ERC) Starting
  Grant. Part of this work done while a DIGITEO visitor in LRI, Orsay.}
}

\begin{document}
\addtocounter{page}{-1}
\maketitle
\thispagestyle{empty}

\begin{abstract}
  We prove an optimal $\Omega(n)$ lower bound on the randomized
  communication complexity of the much-studied
  \textsc{Gap-Hamming-Distance} problem. As a consequence, we
  obtain essentially optimal multi-pass space lower bounds in the
  data stream model for a number of fundamental problems, including
  the estimation of frequency moments.

  The \textsc{Gap-Hamming-Distance} problem is a communication problem,
  wherein Alice and Bob receive $n$-bit strings $x$ and $y$,
  respectively.  They are promised that the Hamming distance between $x$
  and $y$ is either at least $n/2+\sqrt{n}$ or at most $n/2-\sqrt{n}$,
  and their goal is to decide which of these is the case. Since the
  formal presentation of the problem by Indyk and Woodruff (FOCS, 2003),
  it had been conjectured that the na\"ive protocol, which uses $n$ bits
  of communication, is asymptotically optimal.  The conjecture was
  shown to be true in several special cases, e.g., when the
  communication is deterministic, or when the number of rounds of
  communication is limited.

  The proof of our aforementioned result, which settles this conjecture
  fully, is based on a new geometric statement regarding correlations in
  Gaussian space, related to a result of C.~Borell (1985). To prove this
  geometric statement, we show that random projections of not-too-small
  sets in Gaussian space are close to a mixture of translated normal variables.
\end{abstract}
\newpage


\section{Introduction} \label{sec:intro}

Communication complexity is a much-studied topic in computational complexity,
deriving its importance both from the basic nature of the questions it asks
and the wide range of applications of its results, covering, for instance,
lower bounds on circuit depth (see, e.g.,~\cite{KarchmerW90}) and on query
times for static data structures (see,
e.g.,~\cite{MiltersenNSW98,Patrascu08ds}). In the basic setup, which is all
that concerns us here, each of two players, Alice and Bob, receives a binary
string as input. Their goal is to compute some function of the two strings,
using a protocol that involves exchanging a {\em small} number of bits.  Since
communication complexity is often applied as a lower bound technique, much of
the work in the area attempts to rule out the existence of a nontrivial
protocol. For many functions, this amounts to proving an $\Omega(n)$ lower
bound on the number of bits any successful protocol must exchange, $n$ being
the common length of Alice's and Bob's input strings.  Proofs tend to be
considerably more challenging, and more broadly applicable, when the protocol
is allowed to be {\em randomized} and err with some small constant probability
(such as $1/3$) on each input.

For a detailed coverage of the basics of the field, as well as a number of
applications, we refer the reader to the textbook of Kushilevitz and
Nisan~\cite{KushilevitzNisan-book}.  For the reader's convenience, we review the most basic
notions in Section~\ref{sec:prelim}.

In this paper, we focus specifically on the Gap-Hamming-Distance problem
(henceforth abbreviated as $\ghd$), which was first formally studied by Indyk
and Woodruff~\cite{IndykW03} in the context of proving space lower bounds for
the Distinct Elements problem in the data stream model.  We also consider some
closely related variants of $\ghd$.

\mypar{The Problem and the Main Result}
In the Gap-Hamming-Distance problem $\ghd_{n,t,g}$, Alice and Bob receive
binary strings $x\in\b^n$ and $y\in\b^n$, respectively. They wish to decide
whether $x$ and $y$ are ``close'' or ``far'' in the Hamming sense, with a
certain {\em gap} separating the definitions of ``close'' and ``far.''
Specifically, the players must output $0$ if $\dist(x,y) \le t - g$ and $1$ if
$\dist(x,y) > t + g$, where $\dist$ denotes Hamming distance;
if neither of these holds, they may output either $0$ or
$1$. Clearly, this problem becomes easier as the gap $g$ increases. Of special
interest is the case when $t = n/2$ and $g = \Theta(\sqrt n)$; these
parameters are natural, and as we shall show later using elementary
reductions, understanding the complexity of the problem with these parameters
leads to a complete understanding of the problem for essentially all other gap
sizes and threshold locations.
Furthermore, applications of $\ghd$, such as the ones considered by Indyk and
Woodruff~\cite{IndykW03}, need precisely this natural setting of parameters.
Henceforth, we shall simply write ``$\ghd$'' to denote
$\ghd_{n,n/2,\sqrt{n}}$.

Our main result states, simply, that this problem does not have a nontrivial
protocol. Here is a somewhat informal statement; a fully formal version
appears as Theorem~\ref{thm:main}.
\begin{theorem}[Main Theorem, Informal] \label{thm:main-informal}
  If a randomized protocol solves $\ghd$, then it must communicate a
  total of $\Omega(n)$ bits.
\end{theorem}

In fact, the technique we use to prove this theorem yields the stronger
result that the same $\Omega(n)$ hardness holds even if Alice and Bob
are given uniformly random and independent inputs in $\b^n$. The
cleanness of this ``hard distribution'' is potentially important in
applications. We state this result formally in Theorem~\ref{thm:dist}.

\mypar{Relation to Prior Work}
Theorem~\ref{thm:main-informal} is the logical conclusion of a
moderately long line of research. This was begun in the aforementioned
work of Indyk and Woodruff~\cite{IndykW03}, who showed a linear lower
bound on the communication complexity of a somewhat
artificial variant of $\ghd$ in the {\em one-way} model, i.e., in the
model where the communication is required to consist of just one message from
Alice to Bob. Woodruff~\cite{Woodruff04} soon followed
up with an $\Omega(n)$ bound for $\ghd$ itself, still in the one-way
model; the proof used rather intricate combinatorial constructions and
computations.  Jayram et al.~\cite{JayramKS08} later provided a rather
different and much simpler proof, by a reduction from the \textsc{index}
problem. Their reduction was geometric, in the sense that they exploited
a natural correspondence between Hamming space and Euclidean space; this
correspondence has proved fruitful in further work on the problem,
including this work. Recently, Woodruff~\cite{Woodruff09} and Brody and
Chakrabarti~\cite{BrodyC09} gave direct combinatorial proofs of the
$\Omega(n)$ one-way bound.

All of this work left open an important question: {\em what can be said
about the complexity of $\ghd$ when two-way communication is allowed?}
It has been conjectured, at least since the formalization of the problem
in 2003, that $\Omega(n)$ is still the right answer, i.e., that $\ghd$
has {\em no} nontrivial protocol, irrespective of the communication
pattern.

Until 2009, our understanding of this matter was limited to two ``folklore''
results. Firstly, the {\em deterministic} communication complexity of
$\ghd_{n,n/2,g}$ can be shown to be $\Omega(n)$, even allowing two-way
communication and a gap as large as $g = cn$, for a small enough constant $c$.
This follows by directly demonstrating that its communication matrix contains
no large monochromatic rectangles (see, e.g.,~\cite{Woodruff-thesis}).
Secondly, a simple reduction from \textsc{disjointness} to $\ghd_{n,n/2,g}$
shows that its randomized (two-way) communication complexity is $\Omega(n/g)$;
notice that the corresponding bound for $\ghd$ (where $g = \sqrt n$) is
$\Omega(\sqrt n)$.
Meanwhile, we have an {\em upper} bound of $O(n^2/g^2)$, via the simple (and
one-way) protocol that samples sufficiently many coordinates of $x$ and $y$ to
give the right answer with high probability.  It remained a significant
challenge to improve upon either tradeoff, even for just two rounds of
communication.

Recently, Brody and Chakrabarti~\cite{BrodyC09} made progress on
the conjecture, proving it for randomized protocols with two-way
communication, but only a constant number of rounds of communication.
In fact, they showed that in a $k$-round protocol, at least one message
must have length $n/2^{O(k^2)}$. They achieved this via a round
elimination argument. At a high level, they showed that if the first
message in a $\ghd$ protocol is too short, the work done by the rest of
the messages can be used to solve a ``smaller'' instance of $\ghd$, by
exploiting some combinatorial properties of Hamming space. More
recently, Brody et al.~\cite{BrodyCRVW10} improved the bound to
$\Omega(n/(k^2\log k))$, still using a round elimination argument, but
exploiting geometric properties of Hamming and Euclidean space instead.
We refer the reader to the discussion in~\cite{BrodyCRVW10} for details,
including a comparison of the two arguments.

Our main theorem completes this picture, confirming the main outstanding
conjecture about $\ghd$. Moreover, a straightforward reduction
(Prop.~\ref{prop:gap-change}) yields the more general result that the
randomized complexity of $\ghd_{n,n/2,g}$ is $\Theta(\min\{n, n^2/g^2\})$. Our
lower bound proof is significantly different in approach from all of the
aforementioned ones. We now give a high-level overview.

\mypar{The Technique}
Part of the difficulty in establishing our result is that many of the known
techniques for proving communication complexity lower bounds seem unable to
prove bounds better than $\widetilde{\Omega}(\sqrt n)$.
These include the classic rectangle-based methods of discrepancy and
corruption,\footnote{We
assume that the reader has some familiarity with these basic techniques in
communication complexity, which are discussed in detail in the textbook of
Kushilevitz and Nisan~\cite{KushilevitzNisan-book}. Some authors use terms
like ``one-sided discrepancy'' and ``rectangle bound'' when describing the
technique that we (following Beame et al.~\cite{BeamePSW06}) have termed
``corruption.''} for reasons described below. They also include
certain linear algebraic approaches,
such as the factorization norms method of Linial and
Shraibman~\cite{LinialS07} and the pattern matrix method of
Sherstov~\cite{Sherstov11}, because these methods
lower bound {\em quantum} communication complexity. The trouble is that $\ghd$
does have a constant-error $O(\sqrt n \log n)$ quantum communication protocol,
as can be seen by combining a query complexity upper bound due to Nayak and
Wu~\cite{NayakW99} with a communication-to-query reduction, as in Buhrman et
al.~\cite{BuhrmanCW98} or Razborov~\cite{Razborov02}.

Instead, what does work is a suitable generalization of the corruption
method.  Recall that the standard corruption method proceeds as follows.
First, one observes that every protocol that communicates $c$ bits
induces a partition of the communication matrix into $2^c$ disjoint
near-monochromatic rectangles.  In order to show a lower bound of $c$,
one then needs to prove that any rectangle containing at least a
$2^{-c}$ fraction of the $1$-inputs must also contain (or be
``corrupted'' by) a not-much-smaller fraction of the $0$-inputs (or vice
versa).  In other words, one shows that large near-monochromatic
rectangles do not exist, from which the desired lower bound follows.  It
should be noted that proving such a property could be a challenging
task.  Indeed, this is the main technical contribution of Razborov's
proof of the $\Omega(n)$ lower bound on the randomized communication
complexity of the disjointness problem~\cite{Razborov92}.

This idea appears not to give a lower bound better than $\Omega(\sqrt
n)$ on the randomized communication complexity of $\ghd$ because its
communication matrix {\em does} contain ``annoying'' rectangles that are
both large and near-monochromatic.  This can be seen, e.g., by
considering all inputs $(x,y)$ with $x_i = 0$, $y_i = 1$ for $i \in
\{1,2,\ldots,100\sqrt{n}\}$: the resulting rectangle contains a
$2^{-\Theta(\sqrt{n})}$ fraction of all $1$-inputs (it is large), but
a much smaller fraction of $0$-inputs (it is nearly monochromatic).

Our generalization considers not just $0$-inputs and $1$-inputs, but
also a carefully selected set of ``joker'' inputs, whose corresponding
outputs are immaterial.  Loosely speaking, we show that if a large
rectangle contains many more $1$-inputs than $0$-inputs, then the
fraction of joker inputs it contains must be \emph{even larger} than the
fraction of $1$-inputs it contains (by some constant factor, say $3/2$).
This property --- call it the ``joker property'' --- implies that even
though annoying rectangles exist, their union cannot contain more than a
constant fraction of the $1$-inputs (say, $2/3$). In particular, there
is no way to partition the $1$-inputs into $2^c$ near-monochromatic
rectangles, and a lower bound of $c$ follows.

This simple-sounding idea seems to have considerable power. Indeed, the method
we have presented above can be seen as a special case of the ideas behind the
``smooth rectangle bound'' recently introduced by Klauck~\cite{Klauck10} and
systematized by Jain and Klauck~\cite{JainK10}. Formally, when we prove a
communication lower bound using corruption-with-jokers as above, we are
essentially lower bounding the smooth rectangle bound of the underlying
function. For a careful understanding of this matter, based on linear
programming duality, we refer the reader to Jain and Klauck~\cite{JainK10}.

Of course, there remains the task of proving the joker property referred
to above.  It turns out that the statement we need boils down to roughly
the following: for arbitrary sets $A,B\ceq\b^n$ that are not too small (say,
of size at least $2^{0.99n}$), if $x\in_R A$ and $y\in_R B$, then
$\dist(x,y)$ is not too concentrated around $n/2$; a precise statement
appears as Corollary~\ref{cor:cube}. The proof uses a {\em Gaussian
noise correlation inequality} (Theorem~\ref{thm:mainthmgauss}, proved
using analytic methods); this inequality and its proof are the main
technical contributions of the paper and should be of independent
interest.

\mypar{Data Stream and Other Consequences}
The original motivation for studying $\ghd$ was a specific application
to the Distinct Elements problem on data streams. Specifically, given a
stream (sequence) of $m$ elements, each from $[n] := \{1,2,\ldots,n\}$,
we wish to estimate, to within a $1\pm\eps$ factor, the number of
distinct elements in it, while using space sublinear in $m$ and $n$. A
long line of research has culminated in a randomized
algorithm~\cite{KaneNW10} that computes such an estimate (failing
with probability at most $\frac13$, say) in one pass over the stream,
using $O(\eps^{-2} + \log(mn))$ bits of space. 
A space lower bound of $\Omega(\log n)$ has been known for
a while~\cite{AlonMS99} and is easily seen to apply to {\em multi}-pass algorithms.
But the dependence of the lower bound on $\eps$ is a longer story.

An easy reduction (implicit in Indyk and Woodruff~\cite{IndykW03}) shows
that a lower bound of $\Omega(\phi(n,k))$ on the maximum message length
of a $(2k-1)$-round protocol for $\ghd$ would imply a
$\Omega(\phi(\eps^{-2},k))$ space lower bound on $k$-pass algorithms for
the Distinct Elements problem. Thus, the one-way $\Omega(n)$ lower bound
for $\ghd$ implied a tight $\Omega(\eps^{-2})$ lower bound for one-pass
streaming algorithms. The results of Brody and
Chakrabarti~\cite{BrodyC09} and Brody et al.~\cite{BrodyCRVW10} extended
this to $p$-pass algorithms, giving lower bounds of
$\Omega(\eps^{-2}/2^{O(p^2)})$ and $\Omega(\eps^{-2}/(p^2\log p))$,
respectively.

Our main result improves this pass/space tradeoff, giving a space lower
bound of $\Omega(\eps^{-2}/p)$. As is easy to see, this is tight up to
factors logarithmic in $m$ and $n$.  Further, since the communication
lower bound for $\ghd$ can be shown to hold under a uniform input
distribution, this space lower bound can be shown to hold even for
rather benign models of random uncorrelated data~\cite{Woodruff09}.

Suitable reductions from $\ghd$ imply similar space lower
bounds for several other data stream problems, such as estimating frequency
moments~\cite{Woodruff04} and empirical
entropy~\cite{ChakrabartiCM10}. 
One can also derive appropriate lower bounds
for a certain class of distributed computing problems known as functional
monitoring~\cite{ArackaparambilBC09}.
We note that the second
frequency moment (equivalently, the Euclidean norm) can be interpreted
as the self-join size of a table in a database, and
is an especially important primitive needed in many numerical streaming tasks such as 
regression and low-rank approximation.
  %

\mypar{Subsequent Developments}

Since the preliminary announcement of our results~\cite{ChakrabartiR10}, there has been much
additional research related to $\ghd$. One line of research has provided
alternative proofs of our main result.  Vidick~\cite{Vidick11} gave a
proof that followed the same overall outline as ours, but had an
alternative proof of the joker property, based on matrix-analytic and
second moment methods.  More recently, Sherstov~\cite{Sherstov11ghd} gave a
proof that changed the outline itself, working with a closely related
problem called \textsc{gap-orthogonality} that has the advantage of
being amenable to the basic corruption method. Further, by using
an inequality due to Talagrand, Sherstov was able to work with the discrete
problem directly rather than passing to Gaussian space.

Other lines of research have applied the optimal $\Omega(n)$ bound on
the communication complexity of $\ghd$ to obtain results on a diverse
array of topics, including differential privacy~\cite{McGregorMPRTV10},
distributed functional monitoring~\cite{WoodruffZ11}, property
testing~\cite{BlaisBM11}, and data aggregation in
networks~\cite{KuhnO11}. Furthermore, Woodruff and Zhang~\cite{WoodruffZ11} 
have given a new proof of optimal multi-pass space lower bounds for 
Distinct Elements without appealing to our lower bound for $\ghd$.


\section{Corruption, a Generalization, and the Main Theorem}
\label{sec:prelim}

\subsection{Preliminaries}

Consider a communication problem given by a (possibly partial) function
$f:X\times Y \to \{0,1,\star\}$; we let $f$ take the value ``$\star$'' at
inputs for which we do not care about the output given.  For a communication
protocol, $P$, involving two players, Alice and Bob, we write $P(x,y)$ to
denote the output of $P$ when Alice receives $x\in X$ and Bob receives $y\in
Y$. If $P$ is randomized, this is a random variable. We say that $P$ computes
$f$ with error at most $\eps$ if
\[
  \forall\,(x,y)\in X\times Y:~
  f(x,y) \ne \star \Rightarrow \Pr[P(x,y) \ne f(x,y)] \le \eps \, .
\]
When the function $f$ is understood from the context, we use $\err(P)$ to
denote $\inf\{\eps:\,P$ computes $f$ with error at most $\eps\}$. For a
deterministic protocol $P$ and a distribution $\mu$ on $X\times Y$, we
define
\[
  \err_\mu(P) ~:=~ \Pr_{(x,y)\sim\mu}[
    f(x,y) \ne \star ~\wedge~ P(x,y) \ne f(x,y)] \, .
\]

For a protocol $P$, let $\cost(P)$ denote the worst-case number of bits
communicated by $P$. We let $\R_\eps(f)$ and $\D_{\mu,\eps}(f)$ denote the
$\eps$-error randomized and $\eps$-error $\mu$-distributional communication
complexities of $f$, respectively; i.e.,
\begin{align*}
  \R_\eps(f) &~=~ \min \{\cost(P):\, P \text{~is a randomized protocol for~} f
    \text{~with~} \err(P) \le \eps\} \, ; \\
  \D_{\mu,\eps}(f) &~=~ \min \{\cost(P):\, P \text{~is a deterministic protocol for~}
    f \text{~with~} \err_\mu(P) \le \eps\} \, .
\end{align*}
We also put $\R(f) = \R_{1/3}(f)$ and $\D_\mu(f) = \D_{\mu,1/3}(f)$.

\subsection{Rectangles and Corruption}

Consider a two-player communication problem given by a function $f:X\times
Y\to Z$. A set $R \ceq X \times Y$ is said to be a rectangle if $R = \cX
\times \cY$ for some $\cX \ceq X$ and $\cY \ceq Y$. A fundamental property of
communication protocols is the following.
\begin{fact}[Rectangle property; see, e.g.,~\cite{KushilevitzNisan-book}]
\label{fact:rect}
  Let $P$ be a deterministic communication protocol that takes inputs in
  $X\times Y$, produces an output in $Z$, and communicates $c$ bits. Then, for
  all $z\in Z$, there exist $2^c$ pairwise disjoint rectangles
  $R_{1,z},\ldots,R_{2^c,z}$ such that
  \[
    \textstyle \forall\, (x,y) \in X\times Y:~
    P(x,y) = z ~\Longleftrightarrow~ (x,y) \in \bigcup_{i=1}^{2^c} R_{i,z} \, .
  \]
  The rectangles $R_{1,z},\ldots,R_{2^c,z}$ are called the $z$-rectangles of
  $P$.
\end{fact}

Let us focus on problems with Boolean output, i.e., $Z = \b$. The {\em
discrepancy method} for proving lower bounds on $\R(f)$ consists of choosing a
suitable distribution $\mu$ on $X\times Y$ and showing that for every
rectangle $R$, the quantity $|\mu(R \cap f^{-1}(0)) - \mu(R \cap
f^{-1}(1))|$ is ``exponentially'' small.  For several functions, this method is
unable to prove a strong enough lower bound; the canonical example is $\disj$.
A generalization that handles $\disj$, and several other functions, is the
{\em corruption method}~\cite{Razborov92,Klauck03,BeamePSW06} which consists of showing, instead, that for
every ``large'' rectangle $R$, we have $\alpha
\mu_1(R) \le \mu_0(R)$, for a constant $\alpha>0$, where $\mu_i$ is a
probability distribution on $R \cap f^{-1}(i)$, for $i\in\b$. Intuitively, we
are arguing that any large rectangle that contains many $1$s must be corrupted
by the presence of many $0$s. The largeness of $R$ is often
enforced indirectly by writing the inequality in the following manner, where
$m$ typically grows with $|X|$ and $|Y|$:
\begin{equation} \label{eq:corruption-basic}
  \exists\,\alpha_0,\alpha_1> 0~~ \forall\,R~\text{rectangular}:~
  \alpha_1 \mu_1(R) ~\le~ \alpha_0 \mu_0(R) + 2^{-m} \, .
\end{equation}
An inequality of this form allows us to conclude an $\Omega(m)$ lower
bound on $\D_{\nu,\eps}(f)$ for a suitable distribution $\nu$ and
sufficiently small error $\eps > 0$. (Rather than present a full proof, we
note that this follows as a special case of
Theorem~\ref{thm:corruption-joker}, below.) By the easy direction of Yao's
lemma, this implies $\R_\eps(f) = \Omega(m)$.

\subsection{Corruption With Jokers, and the Smooth Rectangle Bound}\label{ssec:corruptionjokers}

We now introduce a suitable generalization of the corruption method,
which, as we shall soon see, implies that $\D_{\mu,\eps}(\ghd) =
\Omega(n)$, for suitable $\mu$ and $\eps$.  The corresponding technical
challenge is met using a new Gaussian noise correlation inequality that
we prove in Section~\ref{sec:gauss-noise}.  Our generalization can be
captured within the very recent {\em smooth rectangle bound}
framework~\cite{Klauck10,JainK10}.  However, we believe that there is
merit in singling out the method we use, because it appears wieldier
than the smooth rectangle bound, which is more technically involved.

The key idea is that, in addition to the distributions $\mu_0$ and
$\mu_1$ on the $0$-inputs and $1$-inputs to $f$, we consider an
auxiliary distribution $\mu_+$ on ``joker'' inputs. Strictly speaking,
we just have a ``joker distribution'' $\mu_+$,\footnote{In the sequel,
when we apply the technique to $\ghd$, $\mu_0$, $\mu_1$ and $\mu_+$ will
be sharply concentrated on pairwise disjoint sets of inputs, which we
can think of as the interesting $0$-inputs, the interesting $1$-inputs,
and the joker inputs, respectively.} and it does not matter how $\mu_+$
relates to $\mu_0$ and $\mu_1$, but it is crucial that the inequality
below gives a negative weight to $\mu_+$, and is therefore a weakening
of~\eqref{eq:corruption-basic}.
%
\begin{equation} \label{eq:corruption-joker}
  \alpha_1 \mu_1(R) - \alpha_+ \mu_+(R)
  ~\le~ \alpha_0 \mu_0(R) + 2^{-m} \, .
\end{equation}

We shall in fact allow a little flexibility in our choice of $\mu_0$ and
$\mu_1$ by requiring only that these be supported ``mostly'' on $0$-inputs and
$1$-inputs. Also, we shall extend our theory to partial functions, since
$\ghd$ is one. The next theorem captures our lower bound technique.

\begin{theorem} \label{thm:corruption-joker}
  For all $\alpha_0,\alpha_1,\alpha_+,\eps > 0$ such that
  $\eps < (\alpha_1 - \alpha_+)/(\alpha_0 + \alpha_1)$,
  there exist $\beta\in\RR$ and $\eps' > 0$ such that the following holds.
  Let $f:X\times Y\to\bstar$ be a partial function.
  Let $A_0 = f^{-1}(0)$ and $A_1 = f^{-1}(1)$.  Suppose that there exist
  distributions $\mu_0,\mu_1,\mu_+$ on $X\times Y$, and a real number
  $m > 0$ such that
  \begin{enumerate}
    \item for $i\in\b$, $\mu_i$ is mostly supported on $A_i$, i.e.,
    $\mu_i(A_i) \ge 1 - \eps$, and
    \item inequality~\eqref{eq:corruption-joker} holds for all
    rectangles $R \ceq X\times Y$.
  \end{enumerate}
  Then, for the distribution
  $\nu := (\alpha_0 \mu_0 + \alpha_1 \mu_1) / (\alpha_0 + \alpha_1)$,
  we have $\D_{\nu,\eps'}(f) \ge m + \beta$.
  In particular, we have $\R_{\eps'}(f) \ge m + \beta$.
\end{theorem}
\begin{proof}
  Consider a deterministic protocol $P$ that computes $f$ with some error
  $\eps'$ (to be fixed later) under $\nu$, and uses $c$ bits of
  communication. Let $R_1,\ldots,R_{2^c}\ceq X\times Y$ be the disjoint
  $1$-rectangles of $P$, as given by Fact~\ref{fact:rect}. Let $S_1 =
  \bigcup_{i=1}^{2^c} R_i$ and $S_0 = X\times Y \setminus S_1$. Notice that
  $S_i$ is exactly the set of inputs on which $P$ outputs $i$. Thus, for
  $i\in\b$, we have
  \begin{align}
    \err_{\mu_i}(P)
    & ~ = ~ \mu_i(S_i \cap A_{1-i}) + \mu_i(S_{1-i} \cap A_i) \notag \\
    & ~\ge~ \mu_i(S_{1-i} \cap A_i) \notag \\
    & ~\ge~ \mu_i(S_{1-i}) - \eps \,, \label{eq:mu-bound}
  \end{align}
  where the last step uses Condition~(1).

  Instantiating inequality~\eqref{eq:corruption-joker} with each $R_i$ and
  summing the resulting inequalities, we get
  \begin{equation} \label{eq:corruption-joker-summed}
    \alpha_1 \mu_1(S_1) - \alpha_+ \mu_+(S_1)
    ~\le~ \alpha_0 \mu_0(S_1) + 2^c\cdot 2^{-m} \, .
  \end{equation}
  Noting that $\mu_1(S_1) = 1 - \mu_1(S_0)$ and applying~\eqref{eq:mu-bound}
  to the $\mu_0$ and $\mu_1$ terms in~\eqref{eq:corruption-joker-summed}, we
  obtain
  \[
    \alpha_1 (1 - \err_{\mu_1}(P) - \eps) - \alpha_+ \mu_+(S_1)
    ~\le~ \alpha_0 (\err_{\mu_0}(P) + \eps) + 2^{c - m} \, .
  \]
  Further, noting that $\mu_+(S_1) \le 1$, and rearranging terms, we obtain
  \begin{align*}
    \alpha_1 - \alpha_+
    & ~\le~ (\alpha_0 + \alpha_1) \eps
        + (\alpha_0 \cdot \err_{\mu_0}(P) + \alpha_1 \cdot \err_{\mu_1}(P))
        + 2^{c - m} \\
    & ~ = ~ (\alpha_0 + \alpha_1) \eps
        + (\alpha_0 + \alpha_1) \err_{\nu}(P) + 2^{c - m} \, .
  \end{align*}
  Using $\err_\nu(P) \le \eps'$ and rearranging further, we get
  \[
    2^{c - m}
    ~\ge~ \alpha_1 - \alpha_+ - (\alpha_0 + \alpha_1)(\eps + \eps') \, .
  \]
  By virtue of the upper bound on $\eps$,
  we may choose $\eps'$ small enough to make the
  right-hand side of the above inequality positive, and equal to $2^\beta$,
  say. Doing so gives us $c \ge m + \beta$, as desired.

  Notice that the ``hard distribution'' $\nu$ is explicitly specified,
  once the distributions involved in Condition~(2) are made explicit.
\end{proof}

We could, alternately, have proved Theorem~\ref{thm:corruption-joker} by
demonstrating that the given conditions imply that the smooth rectangle bound
of $f$ is $\Omega(m)$. We have chosen to give the above proof instead, because
it is more elementary, avoiding the technical details of the latter bound, and
because it was discovered independently by the first named author.

\subsection{Application to GHD: the Main Theorem} \label{ssec:mainthm}

The Gap-Hamming-Distance problem is formalized as the computation of the
partial function $\ghd_{n,t,g}:\b^n\times\b^n\to\{0,1,\star\}$ defined as
follows.
\[
  \ghd_{n,t,g}(x,y) ~=~ \begin{cases}
    0 \, ,     & \> \text{~if~} \dist(x,y) \le t - g \, , \\
    1 \, ,     & \> \text{~if~} \dist(x,y) > t + g \, , \\
    \star \, , & \> \text{~otherwise.}
  \end{cases}
\]
It will be useful to have some flexibility in the choice of the
location of the threshold, $t$, and the size of the gap, $g$. It is
not hard to see that all settings with $t \in \Omega(n) \cap (n - \Omega(n))$
and $g = \Theta(\sqrt n)$ lead to ``equally hard'' problems,
asymptotically; we prove this formally in Lemma~\ref{lem:ghd-shift}.

Rather than working with $\ghd_{n,n/2,\sqrt n}$ directly, it proves
convenient to consider the partial function $f_b = \ghd_{n,n/2-b\sqrt
n,\sqrt{2n}}$, for some large enough constant $b$ to be determined later.
We shall now come up with distributions and constants that satisfy the
conditions of Theorem~\ref{thm:corruption-joker}: Condition~(1)
turns out to be easy to verify, and verifying Condition~(2), as mentioned
above, is a significant technical challenge that we deal with in
Section~\ref{sec:gauss-noise}.

\begin{definition} \label{def:corr-hamming}
  For $p\in[-1,1]$, let $\xi_p$ denote the distribution of
  $(x,y)\in\b^n\times\b^n$ defined by the following randomized procedure: pick
  $x\in_R\b^n$ uniformly at random, and then pick $y$ by independently flipping
  each bit of $x$ with probability $(1-p)/2$. Notice that $\xi_0$ is the
  uniform distribution on $\b^n\times\b^n$.
\end{definition}

We shall need the following two lemmas. The first of these follows easily from
standard tail estimates for the binomial distribution, or even just the Chebyshev bound; we omit its proof. The
second is formally proved at the end of Section~\ref{sec:gauss-noise}.

\begin{lemma} \label{lem:binom-tail}
  For all $\eps > 0$ there exists $b > 0$ such that, for large enough $n$,
  we have
  \begin{align*}
    \xi_{4b/\sqrt n}(A_0)
    & ~ = ~ \Pr_{(x,y)\sim\xi_{4b/\sqrt n}}\left[
      \dist(x,y) \le \frac{n}{2} - (b+\sqrt2)\sqrt n\right] ~\ge~ 1 - \eps \, ,
      \text{~and} \\
    \xi_0(A_1)
    & ~ = ~ \Pr_{(x,y)\sim\xi_0}\left[
      \dist(x,y) \ge \frac{n}{2} - (b-\sqrt2)\sqrt n\right] ~\ge~ 1 - \eps \, ,
  \end{align*}
  where $A_0 = f_b^{-1}(0)$ and $A_1 = f_b^{-1}(1)$.
  \qed
\end{lemma}

\begin{lemma} \label{lem:bool-corr-ineq}
  For all $b > 0$ there exists $\delta > 0$ such that, for large enough $n$,
  we have
  \[
    \forall\,R\ceq\b^n\times\b^n~\text{rectangular}:~~ \textstyle
    \frac12\left(\xi_{-4b/\sqrt n}(R) + \xi_{4b/\sqrt n}(R)\right)
    ~\ge~ \frac23\xi_0(R) - 2^{-\delta n} \, .
  \]
\end{lemma}

To derive the lower bound on $\R(\ghd)$, we put $m = \delta n$, $\mu_0 =
\xi_{4b/\sqrt n}$, $\mu_1 = \xi_0$, $\mu_+ = \xi_{-4b/\sqrt n}$,
$\eps = \frac18$, $\alpha_1 = \frac23$, and $\alpha_0 = \alpha_+ = \frac12$.
Note that this choice of constants satisfies
$\eps < (\alpha_1 - \alpha_+)/(\alpha_0 + \alpha_1)$.
By Lemmas~\ref{lem:binom-tail}
and~\ref{lem:bool-corr-ineq}, we see that Conditions~(1) and~(2),
respectively, of Theorem~\ref{thm:corruption-joker} are met; the inequality in
Lemma~\ref{lem:bool-corr-ineq} is easily seen to be the corresponding
instantiation of~\eqref{eq:corruption-joker}.

Thus, applying Theorem~\ref{thm:corruption-joker}, we conclude that there
exist absolute constants $\eps',\delta,b > 0$ and $\beta\in\RR$ such that, for
large enough $n$, we have $\R_{\eps'}(f_b) \ge \delta n + \beta$.
Combining this with Lemma~\ref{lem:ghd-shift} (proved in Section~\ref{sec:reductions}) to adjust for the
slightly off-center threshold and the size of the gap, and applying standard
error reduction techniques, we obtain the following asymptotically optimal
lower bound for $\ghd$.

\begin{theorem}[Main Theorem] \label{thm:main}
  $\R(\ghd_{n,n/2,\sqrt n}) = \Omega(n)$. \qed
\end{theorem}

In applications of a communication lower bound, it is often helpful to
have a good understanding of the ``hard input distribution'' that
achieves the lower bound.  One slightly unsatisfactory aspect of our
proof above is that the hard distribution for $\ghd$ that it implies is
not too clean. With a little additional work, however, we can show that
the {\em uniform} input distribution is hard for $\ghd$, once we require
a small enough error bound. This is stated in the following theorem,
whose proof appears in Section~\ref{sec:reductions}.

\begin{theorem}[Hardness Under Uniform Distribution] \label{thm:dist}
  There exists an absolute constant $\eps > 0$ for which
  $\D_{\xi_0,\eps}(\ghd_{n,n/2,\sqrt n}) = \Omega(n)$.
\end{theorem}



\def\e{{\rm e}}
\def\d{{\rm d}}
\def\calA{{\cal A}}
\def\calB{{\cal B}}
\def\calC{{\cal C}}
\def\calP{{\cal P}}
\def\calT{{\cal T}}
\def\calH{{\cal H}}
\def\vol{\rm{vol}}
\def\sdp{{\rm{sdp}}}
\def\cc{{\tilde{c}}}
\def\sign{{\rm{sign}}}
\def\id{I}

\def\ra{\rangle}
\def\la{\langle}

\def\bprod{\otimes_b}

\newcommand\set[2]{\left\{ #1 \left|\; #2 \right. \right\}}
\newcommand\sett[1]{\left\{ #1 \right\}}
\newcommand\round[1]{{\lfloor #1 \rceil}}
\newcommand\defeq{\stackrel{def}{=}}
\newcommand\card[1]{\left| #1 \right|}
\newcommand\abs[1]{{\left| {#1} \right|}}
\newcommand\ip[1]{{\left\langle {#1} \right\rangle}}
\newcommand\ipb[1]{{\big\langle {#1} \big\rangle}}
\newcommand\ipB[1]{{\Big\langle {#1} \Big\rangle}}
\newcommand\ipbb[1]{{\bigg\langle {#1} \bigg\rangle}}
\newcommand\norm[1]{{\| #1 \|}}
\newcommand\diver[2]{{D({#1} \,\|\, {#2} )}}

\newcommand\ontop[2]{{
\tiny\begin{array}{c} {#1} \\ {#2} \end{array} }}
\newcommand\ket[1]{{ |{#1} \rangle }}
\newcommand\bra[1]{{ \langle {#1} | }}
\newcommand{\kb}[1]{| #1  \rangle\langle #1|}
\newcommand{\kbbig}[1]{\Bigl| #1  \Bigr\rangle\Bigl\langle #1\Bigr|}

\newcommand{\SAT}{\textsc{SAT}}
\newcommand{\QMS}{\textsc{QMIP$^*$}}
\newcommand{\QMIP}{\textsc{QMIP}}
\newcommand{\Pe}{\textsc{P}}
\newcommand{\NP}{\textsc{NP}}
\newcommand{\QIP}{\textsc{QIP}}
\newcommand{\IP}{\textsc{IP}}
\newcommand{\EXP}{\textsc{EXP}}
\newcommand{\NEXP}{\textsc{NEXP}}
\newcommand{\MIP}{\textsc{MIP}}
\newcommand{\PSPACE}{\textsc{PSPACE}}
\newcommand{\XMIP}{$\oplus$\textsc{MIP}$(2,1)$}
\newcommand{\PCP}{\textsc{PCP}}

\newcommand{\MAXCUT}{\textsc{MaxCut}}
\newcommand{\MAXKCUT}{\textsc{Max-k-Cut}}
\newcommand{\ELIN}{\textsc{E3LIN2}}
\newcommand{\THREESAT}{\textsc{3SAT}}
\newcommand{\SETCOVER}{\textsc{SetCover}}
\newcommand{\MAXCLIQUE}{\textsc{MaxClique}}

\section{An Inequality on Correlation under Gaussian Noise}
\label{sec:gauss-noise}

We now turn to the proof of Lemma~\ref{lem:bool-corr-ineq}, for which we need
some technical machinery that we now develop. We begin with some preliminaries.

\mypar{Some Probability Distributions}
Let $\mu$ denote the uniform (Haar) distribution on $\SS^{n-1}$,
the unit sphere in $\RR^n$. Let $\gamma$ denote the standard Gaussian distribution on $\RR$, with density
function $(2\pi)^{-1/2} \e^{-x^2/2}$, and let $\gamma^n$  denote the $n$-dimensional standard Gaussian
distribution with density $(2\pi)^{-n/2} \e^{-\|x\|^2/2}$.
For a set $A\ceq\RR^n$, when we write, e.g., $\gamma^n(A)$, we tacitly assume that $A$ is measurable. 
For a set $A \subseteq \RR^n$ we denote by $\gamma^n|_A$ the distribution $\gamma^n$ conditioned on
being in $A$. We say that a pair $(x,y)$
is an \emph{$\eta$-correlated Gaussian pair} if its distribution is that
obtained by choosing $x$ from $\gamma^n$ and then setting $y = \eta x + \sqrt{1-\eta^2} z$
where $z$ is an independent sample from $\gamma^n$. It is easy to verify that if $(x,y)$ is
an $\eta$-correlated Gaussian pair, then so is $(y,x)$; in particular, $y$ is
distributed as $\gamma^n$.

\mypar{Relative Entropy}
We recall some basic information theory for continuous probability distributions.
For clarity, we eschew a fully rigorous treatment --- which would introduce a
considerable amount of extra complexity through its formalism --- and instead
refer the interested reader to the textbook of Gray~\cite{Gray-book}.
Given two probability distributions $P$ and $Q$, we define
the \emph{relative entropy} of $P$ with respect to $Q$ as
$$ \diver{P}{Q} ~=~ \int P(x) \ln (P(x)/Q(x)) \, \d x \, .$$
It is well known (and not difficult to show) that the relative entropy is
always nonnegative and is zero iff the two distributions are essentially equal.
We will also need Pinsker's inequality, which says that the statistical distance between
two distributions $P$ and $Q$ is at most $\sqrt{\diver{P}{Q}/2}$
(see, e.g.,~\cite[Lemma~5.2.8]{Gray-book}).
Since we will only consider the relative entropy with respect to the Gaussian
distribution, we introduce the notation
$$ D_{\gamma}(X) ~:=~ \diver{P}{\gamma}$$
where $X$ is a real-valued random variable with distribution $P$. We define $D_{\gamma^n}$ similarly.
These quantities can be thought of as measuring the ``distance from Gaussianity.''
They can be seen, in some precise sense, as additive inverses of entropy, and as such satisfy
many of the familiar properties of entropy. 
For instance, it is easy to verify that
for any sequence of random variables $X_1,\ldots,X_n$ we have the chain rule 
\[
  D_{\gamma^n}(X_1,\ldots,X_n) ~=~ \sum_{k=1}^n D_\gamma(X_k|{X_1,\ldots,X_{k-1}}) \, ,
\]
where, for random variables $X$ and $Y$, we use the notation
$D_{\gamma}(X|Y)$ to denote the expectation over $Y$ of the distance from Gaussianity
of $X|Y$.

\subsection{Projections of Sets in Gaussian Space}

Our main technical result is a statement about the projections of sets in Gaussian space.
More precisely, let $A \subseteq \RR^n$ be any set of not too small measure, say, $\gamma^n(A) \ge \exp(-\delta n)$
for some constant $\delta > 0$.
What can we say about the projections (or one-dimensional marginals) of $\gamma^n|_A$, i.e., the set of distributions of
$\ip{\gamma^n|_A,y}$ as the (fixed) vector $y$ ranges over the unit sphere $\SS^{n-1}$?

Related questions have appeared in the literature.
The first is in work by Sudakov~\cite{Sudakov78} and Diaconis and Freedman~\cite{DiaconisF84}
(see also~\cite{Bobkov03} for a more recent exposition) who showed that for any random variable in $\RR^n$ with zero mean
and identity covariance matrix whose norm is concentrated around $\sqrt{n}$,
almost all its projections are close to the standard normal distribution.
A second related result is by Klartag~\cite{Klartag07} who, building on the previous result but with
considerable additional work, showed that almost all
projections of the uniform distribution over a (properly normalized) convex body are
close to the standard normal distribution. (For the special case of the cube $[-1,1]^n$, this
essentially follows from the central limit theorem.)

Our setting is different as we do not put any restrictions on the set $A$ (such as
convexity) apart from its measure not being too small (and clearly without any
requirement on the measure one cannot say anything about its projections).
Another important difference is that in our setting the projections are \emph{not} necessarily
normal. To see why, take $A=\{x:\, |x_1| > t\}$ for $t \approx \sqrt{\delta n}$, a set with Gaussian measure
roughly $\exp(- \delta n)$, half of which is on vectors with $x_1 \approx t$ and
the other half on vectors with $x_1 \approx -t$.
It follows that the projection of $\gamma^n|_A$ on a unit vector $y$ is distributed more or less like
the mixture of two normal variables, one centered around $t y_1$ and the other centered around $-t y_1$,
both with variance $1$.
For unit vectors $y$ with $|y_1| \ge 1/\sqrt{\delta n}$ (a set of measure about $\exp(-1/\delta)$),
this distribution is very far from any normal distribution.

Our main theorem below shows that the general situation is similar:
for any set $A$ of not too small measure, almost all projections of $\gamma^n|_A$ are close to being mixtures of
translated normal variables of variance $1$. One implication of this (which is essentially all we will use later) is
that for any $A \subseteq \RR^n$ of not too small measure, and $B \subseteq \SS^{n-1}$ whose measure is
also not too small, the inner product $\ip{x,y}$ for $x$ chosen from $\gamma^n|_A$ and $y$ chosen uniformly from $B$ is not
too concentrated around $0$; in fact, it must be at least as ``spread out" as $\gamma$ (and possibly much more).

\begin{theorem}\label{thm:projectionofset}
For all $\eps,\delta>0$ and large enough $n$, the following holds. Let $A \subseteq \RR^n$
be such that $\gamma^n(A) \ge \e^{-\eps^2 n}$. Then, for all but an $\e^{-\delta n/36}$
measure of unit vectors $y \in \SS^{n-1}$, the
distribution of $\ip{x,y}$ where $x \sim \gamma^n|_A$ is equal to the distribution
of $\alpha X+Y$ for some $1-\delta \le \alpha \le 1$ and random variables $X$ and $Y$ satisfying
$$ D_\gamma(X|Y) ~\le~ \eps \, .$$
\end{theorem}

The proof is based on the following two lemmas. The first one below shows that
for any set $A$ whose measure is not too small, and any orthonormal basis, most of the
projections of $\gamma^n|_A$ on the basis vectors are close to normal.
In fact, the statement is somewhat stronger, as it allows us
to condition on previous projections (and this will be crucially used).

\begin{lemma}\label{lem:projections}
For all $\eps>0$ and large enough $n$ the following holds.
For all sets $A \subseteq \RR^n$ with $\gamma^n(A) \ge \e^{-\eps^2 n}$ and all orthonormal bases $y_1,\ldots,y_n$,
at least a $1-\eps$ fraction of the indices $k \in [n]$ satisfy
$$ D_\gamma(P_k|{P_1,\ldots,P_{k-1}}) ~\le~ \eps \, , $$
where $P_i = \ip{u,y_i}$, with $u \sim \gamma^n|_A$.
\end{lemma}
\begin{proof}
By definition,
$ D_{\gamma^n}(\gamma^n|_A) = - \ln \gamma^n(A) \le \eps^2 n$.
Thus, since $(P_1,\ldots,P_n)$ is the vector $u$ written in the orthonormal basis
$y_1,\ldots,y_n$, using the chain rule for relative entropy, we have
\[
  \eps^2 n ~\ge~ D_{\gamma^n}(\gamma^n|_A)
  ~=~ D_{\gamma^n}(P_1,\ldots,P_n)
  ~=~ \sum_{k=1}^n D_\gamma(P_k|{P_1,\ldots,P_{k-1}}) \, .
\]
Hence, for at least a $1-\eps$ fraction of indices $k$, we have
$
  D_\gamma(P_k|{P_1,\ldots,P_{k-1}}) \le \eps.
$
\end{proof}

The second lemma is due to Raz~\cite{Raz99} and shows that any not-too-small subset $B$ of the sphere
contains $n/2$ ``nearly orthogonal'' vectors.
The idea of Raz's proof is the following. First, a simple averaging
argument shows that there is a not-too-small measure of vectors $y'
\in \SS^{n-1}$ satisfying the property that the measure of $B$ inside the unit sphere
formed by the intersection of $\SS^{n-1}$ and the subspace orthogonal to
$y'$ is not much smaller than $\mu(B)$. Second, by the isoperimetric
inequality, almost all vectors in $\SS^{n-1}$ are within distance
$\delta$ of $B$. Together, we obtain a  vector $y'$ as above that is
within distance $\delta$ of $B$. We take $y_{n/2}$ to be the closest
vector in $B$ to $y'$ and repeat the argument recursively with the
intersection of $B$ and the subspace orthogonal to $y'$.

\begin{definition}
A sequence of unit vectors $y_1,\ldots,y_k \in \SS^{n-1}$ is called \emph{$\delta$-orthogonal} if for all $i \in [k]$,
the squared norm of the projection of $y_i$ on $\spn(y_1,\ldots,y_{i-1})$ is at most $\delta$.
\end{definition}

\begin{lemma}[{\cite[Lemma 4.4]{Raz99}}]\label{lem:raz}
For all $\delta>0$ and large enough $n$, the following holds.
Every $B \subseteq \SS^{n-1}$ of Haar measure $\mu(B) \ge \e^{-\delta n /36}$
contains a $\delta$-orthogonal sequence $y_1,\ldots,y_{n/2} \in B$.
\end{lemma}

\begin{proof}[Proof of Theorem~\ref{thm:projectionofset}]
Let $B \subseteq \SS^{n-1}$ be an arbitrary set of unit vectors of measure
at least $\e^{-\delta n/36}$. We will prove the theorem by showing that at
least one vector $y \in B$ satisfies the condition stated in the theorem.

By Lemma~\ref{lem:raz}, there is a sequence of $n/2$ vectors
$y_1,\ldots,y_{n/2} \in B$ that is $\delta$-orthogonal.
Let $y^*_1,\ldots,y^*_{n/2}$ be their Gram-Schmidt orthogonalization, i.e., each $y^*_k$  is
defined to be the projection of $y_k$ on the space orthogonal to $\spn(y_1,\ldots,y_{k-1})$.
Notice that, by definition, we can write each $y_k$ as
$$y_k ~=~ y^*_k + \sum_{i=1}^{k-1} \alpha_{k,i} y^*_i$$
for some real coefficients $\alpha_{k,i}$. Moreover, by assumption,
$\|y^*_k\|^2 \ge 1-\delta$.

Let $P_1,\ldots,P_{n/2}$ be the random variables representing
$\ip{x,y_1^*/\|y^*_1\|},\ldots,\ip{x,y_{n/2}^*/\|y^*_{n/2}\|}$ when $x$ is chosen from $\gamma^n|_{A}$.
By applying Lemma~\ref{lem:projections} to any completion of $y_1^*/\|y^*_1\|,\ldots,y_{n/2}^*/\|y^*_{n/2}\|$ to an orthonormal basis,
we see that there exists an index $k\in [n/2]$ for which
$$ D_\gamma(P_k|{P_1,\ldots,P_{k-1}}) ~\le~ \eps \, .$$
(In fact, at least $1-2\eps$ of the indices $k$ satisfy this.) It remains to notice that we can write
$\ip{x,y_k}$ as
$$\|y^*_k\| P_k + \sum_{i=1}^{k-1} \alpha_{k,i} \|y^*_i\| P_i \, ,$$
which satisfies the condition in the theorem, with $X$ taken to be $P_k$ and
$Y$ taken to be the above sum. Here we are using the fact that $Y$ is a function of
$P_1,\ldots,P_{k-1}$, which implies that $D_\gamma(X|Y) \le
D_\gamma(P_k | P_1,\ldots,P_{k-1})$ since conditioning cannot decrease
relative entropy.
\end{proof}

\subsection{The Correlation Inequality}

We now turn to our main technical result, which is given by the following theorem.

\begin{theorem}\label{thm:mainthmgauss}
For all $c,\eps>0$ there exists a $\delta>0$ such that for all large enough $n$ and $0 \le \eta \le c/\sqrt{n}$ the following holds.
For all sets $A,B \subseteq \RR^n$ with $\gamma^n(A),\gamma^n(B)\ge \e^{-\delta n}$ we have that
$$ \frac{1}{2} \left( \Pr_{(x,y)\text{ is }\eta\text{-correlated}}[x \in A \wedge y \in B]
  + \Pr_{(x,y)\text{ is }-\eta\text{-correlated}}[x \in A \wedge y \in B] \right) ~\ge~ (1-\eps) \gamma^n(A) \gamma^n(B). $$
\end{theorem}

As will become evident in the proof, pairs $(x,y) \in A \times B$ for which $|\ip{x,y}|$ is small contribute
much less to the left hand side than to the right hand side. Hence the theorem essentially
amounts to showing that $\ip{x,y}$ is not too concentrated around zero, and precisely such an anti-concentration
statement is given by Theorem~\ref{thm:projectionofset}.

We point out the following easy corollary (which is in fact
equivalent to Theorem~\ref{thm:mainthmgauss}).

\begin{corollary} \label{cor:centralsymgaussian}
  For all $c,\eps>0$ there exists a $\delta>0$ such that for all large
  enough $n$ and $0 \le \eta \le c/\sqrt{n}$ the following holds.  For
  any sets $A,B \subseteq \RR^n$ with $\gamma^n(A),\gamma^n(B)\ge
  \e^{-\delta n}$ where $A$ (or $B$) is centrally symmetric (i.e.,
  $A=-A$) we have that
  \[
    \Pr_{(x,y)\text{ is }\eta\text{-correlated}}[x \in A \wedge y \in B]
    ~\ge~ (1-\eps) \gamma^n(A) \gamma^n(B) \, .
  \]
\end{corollary}

\begin{remark}
  Without the symmetry assumption, this probability can be considerably
  smaller. For instance, take $A$ and $B$ to be two opposing
  half-spaces, i.e., $A=\{x:\, x_1 < -t\}$ and $B=\{x:\, x_1 > t\}$ for
  $t \approx \sqrt{\delta n}$. Then for $\eta = c/\sqrt{n}$, the
  probability above can be seen to be $\e^{-\Theta(\sqrt{n})} \gamma^n(A)
  \gamma^n(B)$.  In fact, C.~Borell~\cite{Borell85} showed that for any
  given $\gamma^n(A),\gamma^n(B)$ and any $0 \le \eta \le 1$, two opposing
  half-spaces $A,B$ of the corresponding measures exactly achieve the
  minimum of the probability above.  It would be interesting to obtain a
  strengthening of Corollary~\ref{cor:centralsymgaussian} of a similar
  tight nature.  See~\cite{Barthe01} for a short related discussion.
\end{remark}

Recall that $\cosh(x) := \frac{1}{2}(\e^x+\e^{-x})$.
The following technical claim shows that if the distribution of $x$ is close to the normal distribution (in relative
entropy) then the expectation of $\cosh(\alpha x + z)$ is at least $\e^{\alpha^2/2}-\eps$.
Notice that if $x$ is normal, this expectation is
$$\Exp_{x \sim \gamma}[\, \cosh(\alpha x+z)\,] ~=~ \cosh(z) \Exp_{x \sim \gamma}[\,\cosh(\alpha x)\,] ~=~ \cosh(z) \,\e^{\alpha^2/2}  \ge \e^{\alpha^2/2},$$
where in the first equality we used the symmetry of $\gamma$, and the second follows from an easy direct
calculation of the integral (just complete the square in the exponent). 

\begin{claim}\label{clm:coshexp}
For all $\eps,\alpha_0>0$ there exists a $\delta>0$ such that for any probability distribution $P$
on the reals satisfying $D_\gamma(P) < \delta$, any $z \in \RR$, and any $0<\alpha\le\alpha_0$, we have
$$ \Exp_{x \sim P}[\,\cosh(\alpha x + z)\,] ~\ge~ \e^{\alpha^2/2} - \eps.$$
\end{claim}

\begin{proof}
%
Set $M=\Exp_{x \sim \gamma}[\,(1+\cosh(2\alpha_0 x))/\eps \,] $ so that for all $z$ and all $\alpha \le \alpha_0$,
\begin{align*}
\Exp_{x \sim \gamma}[\,\min(\cosh(\alpha x +z), 2M)\,]
&~= \frac{1}{2} \Exp_{x \sim \gamma} [\,  \min(\cosh(\alpha x +z), 2M) + \min(\cosh(\alpha x - z), 2M) \,] \\
&~\ge \Exp_{x \sim \gamma} \left[\, \min\left(\frac{1}{2} ( \cosh(\alpha x +z) + \cosh(\alpha x - z)) , M\right)  \,\right] \\
&~= \Exp_{x \sim \gamma}[\,\min(\cosh(z) \cosh(\alpha x), M)\,] \\
&~\ge \Exp_{x \sim \gamma}[\,\min(\cosh(\alpha x), M)\,] \\
&~\ge \Exp_{x \sim \gamma}[\, \cosh(\alpha x) \,] - \frac{1}{M} \Exp_{x \sim \gamma}[\,\cosh(\alpha x)^2\,] \\
&~= \e^{\alpha^2/2} - \frac{1}{M} \Exp_{x \sim \gamma}\left[\, \frac{1}{2}(1+\cosh(2\alpha x)) \,\right] \\
&~\ge \e^{\alpha^2/2} - \eps/2 \, ,
\end{align*}
where in the third inequality we use the fact that $\min(u,v) \ge u - u^2/v$ for all $u,v>0$.
Next, since
the statistical distance between $P$ and $\gamma$ is at most $\sqrt{2 D_\gamma(P)} < \sqrt{2 \delta}$,
we have that
$$ \Exp_{x \sim P}[\,\cosh(\alpha x + z)\,] ~\ge~
   \Exp_{x \sim P}[\,\min(\cosh(\alpha x + z),2M)\,] ~\ge~
   \e^{\alpha^2/2} - \eps/2 - 2M \sqrt{2 \delta} ~\ge~ \e^{\alpha^2/2} - \eps$$
for small enough $\delta>0$.
\end{proof}

\begin{proof}[Proof of Theorem~\ref{thm:mainthmgauss}]
Let $\beta_1,\beta_2,\beta_3,\beta_4>0$ be small enough constants (depending only on $c$ and $\eps$) to be determined later.
By choosing a small enough $\delta$, and using the concentration of the Gaussian measure around the sphere of radius $\sqrt{n}$ (see, e.g.,~\cite[Lecture 8]{Ball97}), we can guarantee that $A'$, defined as
$$ A' ~=~ \{x\in A:\, (1-\beta_1)n \le \|x\|^2 \le (1+\beta_1)n \} \, ,$$
satisfies $\gamma^n(A') \ge \gamma^n(A) - \beta_2 \e^{-\delta n} \ge (1-\beta_2) \gamma^n(A)$ and similarly
for $B'$. 
We can write
\begin{align*}
 &\Pr_{(x,y)\text{ is }\eta\text{-correlated}}[x \in A \wedge y \in B] \\
 & \qquad \qquad \ge~ \Pr_{(x,y)\text{ is }\eta\text{-correlated}}[x \in A' \wedge y \in B'] \\
  &\qquad \qquad =~ (2\pi)^{-n/2} (2\pi(1-\eta^2))^{-n/2} \int 1_{A'}(x) 1_{B'}(y) \e^{-\|x\|^2/2} \e^{-\|y-\eta x\|^2/2(1-\eta^2)} \d x \d y \\
 & \qquad \qquad =~ (1-\eta^2)^{-n/2} \Exp_{x,y \sim \gamma^n}\big[\,1_{A'}(x) 1_{B'}(y) \e^{-\eta^2 \|x\|^2 / 2(1-\eta^2)} \e^{-\eta^2 \|y\|^2 / 2(1-\eta^2)} \e^{\eta \ip{x,y} / (1-\eta^2)}\,\big] \\
 & \qquad \qquad =~ (1-\eta^2)^{-n/2} \Exp_{x \sim \gamma^n|_{A'},y \sim \gamma^n|_{B'}}\big[\,\e^{-\eta^2 \|x\|^2 / 2(1-\eta^2)} \e^{-\eta^2 \|y\|^2 / 2(1-\eta^2)} \e^{\eta \ip{x,y} / (1-\eta^2)} \,\big] \gamma^n(A') \gamma^n(B') \\
 & \qquad \qquad \ge~ (1-\eta^2)^{-n/2} \,\e^{-\eta^2 (1+\beta_1) n / (1-\eta^2)} \Exp_{x \sim \gamma^n|_{A'},y \sim \gamma^n|_{B'}}\big[\,\e^{\eta \ip{x,y} / (1-\eta^2)} \,\big] \gamma^n(A') \gamma^n(B') \, .
\end{align*}
By averaging this inequality with the analogous one for $-\eta$ and recalling the definition of $\cosh$,
we obtain that the expression we wish to bound is at least
\begin{align}\label{eq:expressiontobound}
  (1-\eta^2)^{-n/2} \,\e^{-\eta^2 (1+\beta_1) n / (1-\eta^2)}  \Exp_{x \sim \gamma^n|_{A'},y \sim \gamma^n|_{B'}}[\,\cosh(\eta \ip{x,y} / (1-\eta^2)) \,] \gamma^n(A') \gamma^n(B') \, .
\end{align}
Let $B'' \subseteq B'$ be the set of all $y \in B'$ for which
$$ \Exp_{x\sim \gamma^n|_{A'}}[\,\cosh(\eta \ip{x,y} / (1-\eta^2)) \,] ~\le~ (1-\beta_3) \e^{(\eta/(1-\eta^2))^2 (1-\beta_1)n /2} \, .$$
We can now complete the proof by showing that $\gamma^n(B'') \le \beta_4 \gamma^n(B')$, since this would imply that
\eqref{eq:expressiontobound} is at least
\begin{align*}
&(1-\eta^2)^{-n/2} \,\e^{-\eta^2 (1+\beta_1) n / (1-\eta^2)} (1-\beta_4) (1-\beta_3) \,\e^{(\eta/(1-\eta^2))^2 (1-\beta_1)n/2} \, \gamma^n(A') \gamma^n(B') \\
&~\ge \e^{n\eta^2/2} \,\e^{-\eta^2 (1+\beta_1) n / (1-\eta^2)} (1-\beta_4) (1-\beta_3) \,\e^{(\eta/(1-\eta^2))^2 (1-\beta_1)n/2} \, (1-\beta_2)^2 \gamma^n(A)\gamma^n(B) \\
&~\ge (1-\eps) \gamma^n(A) \gamma^n(B) \, ,
\end{align*}
assuming $\beta_1$, $\beta_2$, $\beta_3$ and $\beta_4$ are chosen to be sufficiently small and $n$ is large enough.

In order to complete the proof assume, to the contrary, that $\gamma^n(B'') > \beta_4 \gamma^n(B') \ge \beta_4 (1-\beta_2) \e^{-\delta n}$.
Let $\beta_5,\beta_6,\beta_7>0$ be small enough constants to be determined later.
Let $\sqrt{(1-\beta_1)n} \le r \le \sqrt{(1+\beta_1)n} $ be such that
the Haar measure $\mu((r \, \SS^{n-1} \cap B'')/r)$ of points in $B''$ of norm $r$
is at least $\gamma^n(B'')$.
(The existence of such an $r$ follows from the fact that the Gaussian distribution, being spherically symmetric, can be seen as the product
of a certain distribution on radii $r$ and the Haar measure on the sphere. Since $B'' \subseteq B'$, the $r$ maximizing the intersection with the sphere must be in the claimed range.)
We now apply Theorem~\ref{thm:projectionofset} with $\eps$ taken to be $\beta_5$, $\delta$ taken to be $\beta_6$,
and $A$ taken to be $A'$. By taking (our) $\delta$ to be small enough, we obtain a vector $y \in B''$
for which the distribution of $\ip{x,y}$ where $x \sim \gamma^n|_{A'}$ is given by the distribution of
$\alpha r X + r Y$ for some $1-\beta_6 \le \alpha  \le 1 $ and random variables $X$ and $Y$ satisfying
$$ D_\gamma(X|Y)  ~\le~ \beta_5 \, .$$
In particular, we have
$$ \Pr_{Y} [\, D_\gamma(X|Y) \le \sqrt{\beta_5} \,] ~\ge~ 1-\sqrt{\beta_5} \, .$$
Claim~\ref{clm:coshexp} now implies that
\begin{align*}
 \Exp_{x\sim \gamma^n|_{A'}}[\, \cosh(\eta \ip{x,y} / (1-\eta^2)) \, ]
 &~=~ \Exp[\, \cosh(\eta/(1-\eta^2) ( \alpha r X + r Y ))\, ] \\
 &~\ge~ (1-\sqrt{\beta_5}) (\e^{(\eta/(1-\eta^2) \alpha r)^2/2}-\beta_7) \\
 &~\ge~ (1-\sqrt{\beta_5}) (\e^{(\eta/(1-\eta^2))^2  (1-\beta_6)^2 (1-\beta_1)n /2}-\beta_7) \\
 &~>~ (1-\beta_3) \,\e^{(\eta/(1-\eta^2))^2 (1-\beta_1)n /2} \, ,
\end{align*}
assuming $\beta_5$, $\beta_6$ and $\beta_7$ are sufficiently small, in contradiction to
the assumption that $y \in B''$.
\end{proof}

\subsection{Corollary for the Boolean cube}

The Gaussian noise correlation inequality we have just proved implies a
similar statement for the Boolean cube, from which
Lemma~\ref{lem:bool-corr-ineq} follows easily. The statement involves the
distribution $\xi_p$ from Definition~\ref{def:corr-hamming}.

\begin{corollary}[Stronger variant of Lemma~\ref{lem:bool-corr-ineq}] \label{cor:cube}
For all $c,\eps>0$ there exists a $\delta>0$ such that for all large enough $n$ and $0 \le \rrho \le c/\sqrt{n}$ the following holds.
For all sets $A,B \subseteq \{0,1\}^n$ with $|A|,|B|\ge 2^{(1-\delta) n}$, we have that
\[
  \textstyle \frac12\left( \xi_{-\rrho}(A\times B) + \xi_{\rrho}(A\times B) \right)
  ~\ge~ (1-\eps)\, \xi_0(A\times B) \, .
\]
\end{corollary}
To derive Lemma~\ref{lem:bool-corr-ineq}, take $R = A\times B$, $\eps =
\frac13$, and observe that if $\min\{|A|,|B|\} < 2^{(1-\delta)n}$, then
$\xi_0(R) < 2^{-\delta n}$ and the inequality in that lemma holds trivially,
because its right-hand side is negative.

A short calculation shows that the inequality in
Corollary~\ref{cor:cube} is equivalent to
$$ (1-\rrho^2)^{n/2} \, \Exp_{x \in A, y \in B}
   \Big[\, \cosh\Big( \ln\Big(\frac{1+\rrho}{1-\rrho}\Big) \cdot (\dist(x,y)-n/2)  \Big)\, \Big]
   ~\ge~ 1-\eps \, .$$
Hence the corollary can be interpreted as an anti-concentration statement, saying
that for sets $A,B$ that are not too small, the Hamming distance $\dist(x,y)$ between
$x \in_R A$ and $y \in_R B$ cannot be too concentrated around $n/2$.
The quantification is delicate. Notice that already for sets of size $2^{n/2}$ this is no longer the case: take, for instance,
the sets $A = \{0^{n/2}x:\, x\in\b^{n/2}$ and $|x| = n/4\}$ and $B =
\{x0^{n/2}:\, x\in\b^{n/2}$ and $|x| = n/4\}$.

\begin{proof}
Given any $A,B \subseteq \{0,1\}^n$, define
$$ A' ~=~ \{ x \in \RR^n:\, \sign(x) \in A\} $$
where $\sign(x) \in \{0,1\}^n$ is the vector indicating
the sign of each coordinate of $x$, and define $B'$ similarly.
Then it is easy to check that $\gamma^n(A') = |A|/2^n$ and $\gamma^n(B') =
|B|/2^n$, so that $\gamma^n(A')\gamma^n(B') = \xi_0(A \times B)$,
and that for all $\eta$,
\[
  \Pr_{(x,y)\text{ is }\eta\text{-correlated}}[x \in A' \wedge y \in B']
  ~=~ \xi_{\rrho}(A \times B)
\]
for $\rrho = 1-\frac{2}{\pi}\arccos \eta$ (since the probability
that $\sign(x) \neq \sign(y)$ when $x,y \in \RR$ are $\eta$-correlated
can be computed to be $\frac{1}{\pi} \arccos \eta$).
For small $\rrho$, we get
$\rrho \approx \frac{2}{\pi} \eta$, and the corollary follows from
Theorem~\ref{thm:mainthmgauss}.
\end{proof}


\section{Reductions, Related Results and Generalizations} \label{sec:reductions}

Recall that our argument in Section~\ref{ssec:mainthm} gave an
$\Omega(n)$ lower bound on $\R(\ghd_{n,n/2-b\sqrt n,\sqrt{2n}})$, for a
certain constant $b$. To obtain an $\Omega(n)$ bound for $\ghd$ itself
(which, we remind the reader, is shorthand for $\ghd_{n,n/2,\sqrt n}$),
we use a toolkit of simple reductions, given in the next lemma.
Furthermore, using the toolkit, we can generalize the $\ghd$ bound to
cover most parameter settings, and using similarly simple reductions, we
can obtain optimal lower bounds for related problems.

\begin{lemma} \label{lem:scale-center-shift}
  For all integers $n,k,\ell,m$ and reals $t,g,g' \in [0,n]$, with $n,k > 0$
  and $g' \ge g$, the following relations hold.
  \begin{enumerate}
    \item $\R(\ghd_{n,t,g'}) \le \R(\ghd_{n,t,g})$.
    \item $\R(\ghd_{n,t,g}) \le \R(\ghd_{kn,kt,kg})$.
    \item $\R(\ghd_{n,t,g}) \le \R(\ghd_{n+\ell+m,t+\ell,g})$.
    \item $\R(\ghd_{n,t,g}) = \R(\ghd_{n,n-t,g})$.
  \end{enumerate}
\end{lemma}
\begin{proof}
  We give brief sketches of the proofs of these statements.
  \begin{enumerate}
    \item A correct protocol for $\ghd_{n,t,g}$ is also one for
    $\ghd_{n,t,g'}$.
    \item We can solve $\ghd_{n,t,g}$ by having Alice and Bob ``repeat'' their
    $n$-bit input strings $k$ times each --- which has the effect of also
    amplifying the gap by a factor of $k$ --- and then simulating a protocol
    for $\ghd_{kn,kt,kg}$.
    \item We can solve $\ghd_{n,t,g}$ by having Alice pad her input by
    appending the string $0^{\ell+m}$ to it, Bob pad his by appending $1^\ell
    0^m$ to it, and then simulating a protocol for $\ghd_{n+\ell+m,t+\ell,g}$.
    \item Alice flips each bit of her input and the parties then simulate a
    protocol for $\ghd_{n,n-t,g}$. \qedhere
  \end{enumerate}
\end{proof}

As promised, using parts of the above lemma, we establish the following
lemma, which formally completes the proof of the main theorem.

\begin{lemma} \label{lem:ghd-shift}
  For all integers $n > 0$ and reals $b > 0$, with $n/2 \ge b\sqrt n$, we have
  $\R(\ghd_{n,n/2-b\sqrt n,\sqrt{2n}}) \le \R(\ghd_{2n,n,\sqrt{2n}})$.
\end{lemma}
\begin{proof}
  Apply part~(3) of Lemma~\ref{lem:scale-center-shift}, with $\ell = n/2
  + b\sqrt n$ and $m = n/2 - b\sqrt n$.
\end{proof}
The previous lemma can in fact be generalized, by invoking the remaining parts of
Lemma~\ref{lem:scale-center-shift}, to obtain a lower bound that handles all
thresholds $t$ that are not too close to either end of the interval $[0,n]$. We omit
the details, which are routine, if somewhat tedious.
\begin{prop} \label{lem:ghd-shift-generalized}
  For all reals $a \in (0,\frac12]$ and $b > 0$, and all large enough integers
  $n$, the following holds. Let $t,g$ be reals with $t \in [an,(1-a)n]$ and $g
  \le b\sqrt n$. Then $\R(\ghd_{n,t,g}) = \Omega(n)$. \qed
\end{prop}

The next result resolves the randomized complexity of
$\ghd_{n,n/2,g}$ for a general gap size, $g$.

\begin{prop} \label{prop:gap-change}
  For integers $n$ and $g$, with $1 \le g \le n$, we have
  $\R(\ghd_{n,n/2,g}) = \Theta(\min\{n, n^2/g^2\})$.
\end{prop}
\begin{proof}
  For the upper bound, consider the protocol where Alice and Bob, on input
  $(x,y)\in\b^n\times\b^n$, use public randomness to select a subset $S \ceq
  [n]$ uniformly at random, from amongst all subsets of a certain size, $k$,
  compute $d = |\{i\in S:\, x_i \ne y_i\}|$ by brute
  force (say, with Alice sending Bob the bits $x_i$ for $i\in S$), and output
  $0$ if $d \le k/2$ and $1$ if $d > k/2$. This protocol clearly communicates
  $k$ bits, and an easy application of the
  Chernoff bound shows that this gives a $\frac13$-error protocol if we choose
  $k = O(n^2/g^2)$.

  For the lower bound, we may assume that $g > \sqrt n$, for otherwise the
  claim is obviously true. Applying part~(2) of
  Lemma~\ref{lem:scale-center-shift} with $k = g^2/n$ (for simplicity, we
  ignore divisibility issues), we obtain $\R(\ghd_{n^2/g^2,n^2/2g^2,n/g}) \le
  \R(\ghd_{n,n/2,g})$. The result follows by applying Theorem~\ref{thm:main}
  to the left-hand side of this inequality.
  %
\end{proof}

\subsection{Hardness Under Uniform Distribution}

We now turn to proving Theorem~\ref{thm:dist}, which extends the
$\Omega(n)$ lower bound for $\ghd$ to the specific input distribution
$\xi_0$, the uniform distribution on $\b^n\times\b^n$.

\begin{proof}[Proof of Theorem~\ref{thm:dist}]
  For an integer $n$ and real $p\in[-1,1]$, let $\mu_{n,p}$ denote
  the binomial distribution with parameters $n$ and $(1-p)/2$; notice
  that $\mu_{n,0}$ is the symmetric binomial distribution. Let $P$ be a
  deterministic protocol for $\ghd_{2n,n,\sqrt{2n}}$ such that
  $\err_{\mu_{2n,0}}(P) = \delta$.
  Our goal is to show that if $\delta$ is small enough then $\cost(P) = \Omega(n)$.
  For $d \in \{0,1,\ldots,2n\}$, let $\delta_d$ be the error probability of $P$ on uniform inputs at distance $d$, i.e.,
  \[
    \delta_d ~:=~ \Pr_{(x,y)\sim\xi_0}\left[
      \ghd_{2n,n,\sqrt{2n}}(x,y) \ne \star ~\wedge~
      P(x,y) \ne \ghd_{2n,n,\sqrt{2n}}(x,y) \mid \dist(x,y) = d
    \right] \, .
  \]
  Then, we have
  \[
    \delta ~=~ \sum_{d=0}^{2n} \mu_{2n,0}(d) \delta_d \, .
  \]

  Let $Q$ be the following protocol for $\ghd_{n,n/2-b\sqrt
  n,\sqrt{2n}}$. On input $(x,y)$, Alice and Bob first pad their inputs
  as in Lemma~\ref{lem:ghd-shift}. Then, using
  public randomness, they choose $z\in_R \b^{2n}$ and a random
  permutation $\sigma \in_R \mathcal{S}_{2n}$, and then each player adds
  $z$ bitwise to their padded input and permutes the coordinates of the
  result according to $\sigma$. Let $x',y'\in\b^{2n}$ be the parties'
  respective inputs after these transformations.  Alice and Bob solve
  their problem by simulating $P$ on input $(x',y')$.  It is easy to see
  that $(x',y')$ is uniformly distributed among all pairs with Hamming
  distance $n/2 + b\sqrt n + \dist(x,y)$.

  Let $\nu$ denote the hard distribution for $\ghd_{n,n/2-b\sqrt
  n,\sqrt{2n}}$ implied by our proof of Theorem~\ref{thm:main}. To be
  explicit, we have $\nu = \frac37 \xi_{4b/\sqrt n} + \frac47 \xi_0$.
  Let $\lambda := \frac37 \mu_{n,4b/\sqrt n} + \frac47 \mu_{n,0}$ be the
  corresponding distribution of Hamming distances. It then follows that
  \[
    \err_\nu(Q) ~=~ \sum_{d=0}^n \lambda(d) \delta_{d + n/2 + b\sqrt n} \, .
  \]
  Suppose we are given a constant $\alpha > 0$.  From standard
  properties of the binomial distribution, it follows that there exist
  reals $c, K > 0$ (depending on $\alpha$ and $b$, but independent of
  $n$) such that
  \[
    \sum_{d=0}^{n/2-c\sqrt n} \lambda(d) +
    \sum_{d=n/2+c\sqrt n}^n \lambda(d) ~\le~ \alpha \, ,
  \]
  and for integers $d \in [n/2-c\sqrt n, n/2+c\sqrt n]$,
  \[
    \lambda(d) ~\le~ K \mu_{2n,0}(d + n/2 + b\sqrt n) \, .
  \]
  It then follows that $\err_\nu(Q) \le \alpha + K\delta$. By picking
  $\alpha$ sufficiently small, we obtain by our
  proof of Theorem~\ref{thm:main} that, for small enough $\delta$, $\cost(Q) = \Omega(n)$. Since $Q$ communicates exactly as
  many bits as $P$, it follows that $\cost(P) = \Omega(n)$.
\end{proof}

\subsection{Related Communication Problems with a Gap}

We remark that results similar to those for $\ghd$ also hold for {\sc
gap-intersection-size}, where Alice and Bob have sets $x,y\ceq[n]$ as inputs
and are required to distinguish between the cases $|x\cap y| \le t - g$ and
$|x\cap y| > t + g$, for a threshold parameter $t$ and gap size $g$. Let this
problem be denoted by {\sc gis}$_{n,t,g}$.  We then have the following result,
by an easy reduction from $\ghd$.
\begin{prop}
  Suppose $t \in \Omega(n) \cap (n - \Omega(n))$ and $g = \Theta(\sqrt n)$.
  Then $\R(\textsc{gis}_{n,t,g}) = \Omega(n)$.
  \qed
\end{prop}

Finally, we also remark that results similar to those for $\ghd$ also hold for
the closely related (in fact, essentially equivalent) problem {\sc
gap-inner-product}. Here, Alice and Bob have $d$-dimensional unit vectors $x,y$
as inputs and are trying to distinguish between the cases
$\ip{x, y} \ge \eps$ and $\ip{x, y} \le -\eps$. There is a simple $O(1/\eps^2)$
protocol for this problem: the players use shared randomness to choose
$O(1/\eps^2)$ random hyperplanes and then compare which side of each hyperplane their
inputs lie in. Our main theorem implies that this is tight assuming $d \ge 1/\eps^2$,
as can be seen by embedding the hypercube in the set $\{-1/\sqrt{n},1/\sqrt{n}\}^n$.


\section*{Acknowledgments}

We thank Bo'az Klartag for referring us to~\cite{Barthe01}, Thomas
Vidick for comments on an earlier draft and David Woodruff for
encouraging us to include a proof of Theorem~\ref{thm:dist}. Oded Regev thanks
Hartmut Klauck for introducing him to the smooth rectangle bound. Amit
Chakrabarti thanks Oded Regev for agreeing to include a complete proof of the Gaussian
noise correlation inequality in this paper, and T.~S.~Jayram for many
enlightening discussions about $\ghd$, over the years.

{\small
  \bibliographystyle{alphaabbrvprelim}
  \bibliography{ghd}
}


\appendix

\end{document}